\documentclass[12pt]{article}
\usepackage{amsmath}
\usepackage{latexsym}
\usepackage{bbm}

\def\a{\alpha}

\def\b{\beta}
\def\g{\gamma}

\def\d{\delta}

\def\th{\theta}

\def\l{\lambda}
\def\m{\mu}
\def\n{\nu}

\def\r{\rho}

\def\s{\sigma}
\def\t{\tau}

\def\O{\Omega}

\def\pa{\partial}

\def\and{{\rm and}}

\def\ie{{\it i.e.,} }

\def\IZ{\mathbbm Z}

\begin{document}
\vspace*{-1.0in}
\thispagestyle{empty}
\begin{flushright}
CALT-TH-2014-149
\end{flushright}


\vspace{1.0in} {\Large
\begin{center}
GAUGE THEORIES ON THE COULOMB BRANCH
\end{center}}\vspace{.25in}

\begin{center}
John H. Schwarz\footnote{jhs@theory.caltech.edu}
\\
\emph{Walter Burke Institute for Theoretical Physics\\
California Institute of Technology\\ Pasadena, CA  91125, USA}
\end{center}
\vspace{.25in}

\begin{center}
\textbf{Abstract}
\end{center}
\begin{quotation}
\noindent

We construct the world-volume action of a probe D3-brane in
$AdS_5 \times S^5$ with $N$ units of flux. It
has the field content, symmetries, and dualities
of the $U(1)$ factor of ${\cal N} =4$ $U(N+1)$
super Yang--Mills theory, spontaneously broken to $U(N) \times U(1)$ by being
on the Coulomb branch, with the
massive fields integrated out. This motivates the conjecture that it is
the exact effective action, called a
{\em highly effective action} (HEA). We construct
an $SL(2,\IZ)$ multiplet of BPS soliton solutions of the D3-brane theory
(the conjectured HEA) and show that they reproduce the electrically charged
massive states that have been integrated out as well as magnetic monopoles and dyons.
Their charges are uniformly spread on a spherical surface, called {\em a soliton bubble},
which is interpreted as a phase boundary.

\end{quotation}

\vspace{1.0in}
\centerline{Lectures presented at the Erice 2014 International School of Subnuclear Physics}

\newpage

\pagenumbering{arabic}

\tableofcontents

\newpage

\section{Introduction}

Two recent papers \cite{Schwarz:2013wra} \cite{Schwarz:2014rxa} explored the construction
and properties of certain actions that describe superconformal
gauge theories on the Coulomb branch. The examples that were studied are ones with a lot
of supersymmetry. The reason for this choice is that the large symmetry gives us more confidence
in the plausibility of the conjecture that we will describe.

Even though \cite{Schwarz:2013wra} discussed several examples, this article only describes
one of them, namely ${\cal N} =4$ super Yang--Mills (SYM) theory.
Our procedure is to begin by deriving the world-volume action of a probe D3-brane in an
$AdS_{5}\times S^5$ background of type IIB superstring theory.  As will be explained,
the probe D3-brane action involves various approximations. However, the resulting
formula has a number of exact properties, especially symmetries, that make it an
attractive candidate for the exact solution to a different problem. Specifically, we
conjecture that it is the exact
effective action for ${\cal N} =4$ SYM theory on the Coulomb branch, called
a {\em highly effective action} (HEA).

More generally, we consider the world-volume action of a
probe $p$-brane in an $AdS_{p+2} \times M_q$
background. $M_q$ is a $q$-dimensional compact space with $N$ units of
$q$-form flux, $\int_{M_q} F_q \sim N$.
In addition to a D3-brane in $AdS_5 \times S^5$, ref.~\cite{Schwarz:2013wra} also discussed
\begin{itemize}
\item M2-brane in $AdS_4 \times S^7/\IZ_k$
\item D2-brane in $AdS_4 \times CP^3$
\item M5-brane in $AdS_7 \times S^4$
\end{itemize}
The first two of these correspond to ABJM theory in three dimensions, whereas
the third one corresponds to the $(2,0)$ theory in six dimensions.
As our understanding improves, we intend
to extend these investigations to theories with less supersymmetry.

Section 2 provides background material in string theory and quantum field theory
that is required to follow the rest of the story.
People knowledgable in these subjects can skip this
section. Section 3 constructs (the bosonic part of) the formula for the probe
D3-brane action -- the conjectured HEA -- and discusses some of its properties.
Section 4, based on \cite{Schwarz:2014rxa},
describes the construction of BPS soliton solutions of this action. We find exactly
the spectrum of soliton solutions expected for the HEA. Moreover, they
have an interesting structure in which the charge of the soliton resides on a spherical
shell. Even though the theory is nongravitational, the solitons turn out to have
unexpected analogies with black holes. This analogy suggests that it may be
possible to associate an entropy to solitons with a large charge.

\section{Background Material}

\subsection{${\cal N}=4$ super Yang--Mills theory}

When one says that a theory has ${\cal N}$ supersymmetries, one means that the conserved
supercharges consist of ${\cal N}$ irreducible spinors, as appropriate for the spacetime
dimension in question. ${\cal N} =4$ supersymmetry is the maximal amount that is possible for an interacting nongravitational quantum field theory in four-dimensional spacetime (4d).
The four spinors may be chosen
to be two-component complex Weyl spinors. Thus, there are
16 real supercharges. It is believed that the only 4d nongravitational quantum field theories
with this amount of supersymmetry are SYM theories.
${\cal N} =4$ SYM exists for any compact gauge group, but we will focus on the case of $U(N)$,
so all of the fields are $N \times N$ hermitian matrices.

${\cal N} =4$ SYM was originally derived by first constructing SYM theory in 10d \cite{Brink:1976bc}. It contains a ten-vector gauge field $A_M$ and a Majorana--Weyl spinor
$\Psi$, both in the adjoint representation of the gauge group.
The 4d theory was then obtained by dimensional reduction,
which simply means dropping the dependence of the fields on six of the (Euclidean)
spatial dimensions.
This gives a 4d theory containing a four-vector $A_\m$, six scalars $\phi^I$, and four Weyl spinors $\psi^A$. The 10d SYM theory is not a consistent quantum field theory (by itself), but the
4d SYM theory obtained in this way is a consistent quantum theory.
It has an $SU(4) \sim SO(6)$ global symmetry,
which corresponds to rotations of the six extra dimensions.
Because this symmetry also rotates the four supercharges, $Q^A$, it is called an R-symmetry.

${\cal N} =4$ SYM theories are conformally invariant, which implies that
they are ultraviolet (UV) finite.
This was the first class of UV finite 4d quantum field theories to be discovered,
but many more (with less supersymmetry) are now known. Combining all of the spacetime symmetries,
which are Lorentz invariance, translation invariance, supersymmetry, scale invariance,
conformal invariance, R-symmetry, and conformal supersymmetry, gives the superconformal supergroup
called $PSU(2,2|4)$. (A supergroup has both bosonic and fermionic generators.)
Its bosonic subgroup is $SU(2,2) \times SU(4)$. The first factor is the 4d conformal group and the second factor is the R-symmetry. Anticommutators of the supersymmetry charges
give the momentum operators, which are generators of spacetime translations. Similarly, anticommutators of conformal supersymmetry charges give operators that generate conformal transformations. Mixed anticommutators give the rest of the bosonic generators.
Counting both the Poincar\'e and conformal supersymmetries, there are a total of 32
fermionic generators.

The parameters of ${\cal N} =4$ SYM with gauge group $U(N)$ are a YM coupling constant $g$,
a vacuum angle $\th$, and the rank of the gauge group $N$. For large $N$ and fixed
't Hooft parameter
\begin{equation}
\l = g^2 N,
\end{equation}
the theory has a $1/N$ expansion (for large $N$) with a nice topological interpretation.
The leading term in this expansion (the planar approximation)
has an additional symmetry called dual conformal
invariance. This amount of symmetry is sufficient to make the theory
completely integrable. It is not known whether this only applies to the planar approximation
or whether it extends to the complete theory. Defining the complex parameter
\begin{equation}
 \t = \frac{\th}{2\pi} + i \frac{4\pi}{g^2},
\end{equation}
the $U(N)$ theory has an $SL(2,\IZ)$ duality symmetry under which
\begin{equation}
  \t \to \frac{a\t + b}{c\t + d},
\end{equation}
where $a,b,c,d$ are arbitrary integers satisfying $ad-bc=1$.
The transformation $\t \to -1/\t$ is called S duality.
When $\th =0$, this gives $g \to 4\pi/g$, which allows one to relate
strong coupling to weak coupling. S-duality is an exact nonabelian
electric-magnetic equivalence.

\subsection{Type IIB superstring theory}

Type IIB superstring theory is one of the five distinct superstring theories in
10d. The IIB theory has two Majorana--Weyl supersymmetries of the same chirality,
for a total of 32 real conserved supercharges. Its massless bosonic fields are
\begin{itemize}
\item the 10d metric $g_{MN}$
\item the dilaton $\s$
\item the Neveu-Schwarz (NS-NS) two-form $B_{MN}$
\item Ramond-Ramond (RR) zero-, two-, and four-forms $C$, $C_{MN}$, and $C_{MNPQ}$
\end{itemize}
The four-form $C_4$, constructed from $C_{MNPQ}$, has a self-dual field strength
$F_5 = d C_4 + \ldots$.

Type IIB superstring theory has a few solutions that preserve all of the supersymmetry.
The most obvious one is 10d Minkowski spacetime, \ie $g_{MN} = \eta_{MN}$. In this solution
$\s$ and $C$ are constants and the other fields vanish.
A less obvious maximally supersymmetric solution has a metric describing the
geometry $AdS_5\times S^5$ -- we'll give the formula later. In this solution $\s$ and $C$ are
again constants. However, now $F_5 \sim N[{\rm vol}(AdS_5) + {\rm vol}(S^5)]$. $N$ is the number
of units of five-form flux threading the five-sphere, $\int_{S^5} F_5 \sim N$, where the
coefficients depend on normalization conventions.
The isometry of this solution is given by the supergroup $PSU(2,2|4)$, the same supergroup
as before! In particular, $SU(2,2)$ is the isometry of $AdS_5$ and $SU(4)$ is the isometry of $S^5$.

The value of $\exp(\s)$ is the string coupling constant $g_s$ and the
value of $C$ is called $\chi$. In terms of these one can form
\begin{equation}
  \t = \chi + i/{g_s}.
\end{equation}
Type IIB superstring theory has an exact $SL(2,\IZ)$ symmetry under which
\begin{equation}
  \t \to \frac{a\t + b}{c\t + d},
\end{equation}
as before. More generally, the fields $\s$ and $C$ transform in this manner, but we
are concerned with the case when they are constants.

\subsection{AdS/CFT duality}

The preceding sections have been presented so as to make the AdS/CFT conjecture seem obvious.
However, when it is was put forward by Maldacena \cite{Maldacena:1997re}, it came as
quite a surprise. Specifically, he proposed that
${\cal N} =4$ SYM in 4d with $U(N)$ gauge group is exactly equivalent (``dual'')
to type IIB superstring theory in an $AdS_5 \times S^5$ background with $N$ units of five-form flux. There are many other analogous AdS/CFT pairs, some of which are relevant to the other
examples of probe-brane theories listed in the introduction.

The evidence for AdS/CFT duality that we have presented so far is that both the gauge theory
and the string theory have $PSU(2,2|4)$ symmetry and $SL(2, \IZ)$ duality.
However, by now there is much, much more evidence, which we will
not describe here. Even though the evidence is overwhelming,
a complete proof is not possible. The reason for saying this is that we lack a
complete nonperturbative definition of
type IIB superstring theory (other than the one given by AdS/CFT duality). The best one can hope to
do, in the absence of such a definition, is to show that everything we know about type IIB
superstring theory in an $AdS_5 \times S^5$ background with $N$ units of five-form flux agrees
with what can be deduced from ${\cal N} =4$ SYM in 4d with $U(N)$ gauge group.

\subsection{D3-branes}

Superstring theories contain various supersymmetric (and hence stable) extended objects.
Ones with $p$ spatial dimensions are called $p$-branes.
They carry a type of conserved current $J$ that couples to a $(p+1)$-form gauge
field $A$ with a $(p+2)$-form field strength $F=dA$.
Those $p$-branes on which strings can end are called D$p$-branes. (D stands for Dirichlet
here, since such strings satisfy Dirichlet boundary conditions in the directions orthogonal
to the $p$-brane. They also satisfy Neumann boundary conditions in the directions parallel
to the D$p$-brane.)

In type IIB superstring theory supersymmetric D$p$-branes exist for $p=1,3,5,7$.
Only the D3-brane is invariant under $SL(2,\IZ)$ transformations. This theory has an
infinite number of different kinds of strings. A $(p,q)$ string arises as a bound
state of $p$ fundamental strings and $q$ D-strings, provided $p$ and $q$ are relatively
prime. These strings transform irreducibly under $SL(2,\IZ)$, and any one of them can
end on a D3-brane. From the point of view of the world-volume theory of a single D3-brane,
which is an abelian gauge theory, the end of the string appears to carry
$p$ units of electric charge and $q$ units of magnetic charge.

The various $p$-branes also act as sources of gravitational and other fields.
Maldacena was led to his conjecture
by realizing that $N$ coincident D3-branes give a ``black brane,'' which is a
higher-dimensional generalization of a black hole, whose near-horizon geometry is
$AdS_5 \times S^5$ with $N$ units of 5-form flux on the sphere. In this way the
branes are replaced by a 10d geometry with a horizon, and the 10d string
theory is represented ``holographically'' by a 4d quantum field theory.

A stack of coincident flat D$p$-branes has fields that are localized on its
$(p+1)$-dimensional world volume. They define a ``world-volume theory,'' which is maximally supersymmetric when the background in which they are embedded is maximally supersymmetric.
In particular, we will study the world-volume theory of a single D3-brane embedded in
an $AdS_5 \times S^5$ background geometry, which may be regarded as having been created by
$N$ other coincident D3-branes.

\subsection{The Coulomb branch}

A stack of $N$ coincident flat D3-branes has a world-volume theory, which is a $U(N)$ gauge
theory. Consider starting with $N$ coincident flat D3-branes
and pulling them apart along one of the
orthogonal axes (let's call it the $x$ direction) into clumps $N_1 + N_2 + \ldots +N_k = N$.
Then, the world-volume gauge symmetry is broken spontaneously,
\begin{equation}
U(N) \to U(N_1) \times U(N_2) \times \ldots \times U(N_k).
\end{equation}
The world-volume fields that arise as the lowest mode of a fundamental string with one end attached
to the $i$th D3-brane and the other end attached to the $j$th D3-brane acquire the mass
\begin{equation}
m_{ij} = |x_i -x_j| T,
\end{equation}
where $T$ is the fundamental string tension. This is similar to the Higgs mechanism. The main
difference is that the scalar fields that are eaten by the gauge fields that become massive
belong to the adjoint representation of the original gauge group (in contrast to the $SU(2)$
Higgs doublet in the standard model). To emphasize this distinction, this phase of the
gauge theory is called a Coulomb branch (rather than a Higgs branch).

Let us consider ${\cal N} =4$, $d=4$ SYM theory with a $U(N)$ gauge group.
For most purposes one can say that a free $U(1)$ multiplet decouples leaving
an $SU(N)$ theory. However, this `decoupled' $U(1)$ is needed to get the
full $SL(2,\IZ)$ duality group rather than a subgroup \cite{Belov:2004ht}\cite{Aharony:2013hda}.
In the special case of $U(2)$, if we ignore the decoupled $U(1)$, the remaining
$SU(2)$ theory on the Coulomb branch has unbroken $U(1)$ gauge symmetry. Let
us refer to the massless supermultiplet as the ``photon'' supermultiplet and the two massive supermultiplets as $W^{\pm}$.

The $U(2)$ ${\cal N} =4$ SYM theory on the Coulomb branch
has a famous soliton solution: the 't Hooft--Polyakov monopole. This solution
preserves half of the supersymmetry. One says that it is ``half BPS.'' This monopole is part
of an infinite $SL(2,\IZ)$ multiplet of half-BPS
$(p,q)$ states, with $p$ units of electric charge and $q$ units of magnetic charges,
where $p$ and $q$ are coprime.
The masses, determined by supersymmetry, are
$$ M_{p,q} = v g |p+q\t|= vg\sqrt{\left(p + \frac{\th}{2\pi}q\right)^2 + \left( \frac{4\pi q}{g^2}\right)^2}$$
where $\langle\phi\rangle =v $ is the vev of a massless scalar field.
The $W$ mass is $M_{1,0} = gv$ and the monopole mass is $M_{0,1} = 4\pi v/g$ for $\th =0$.
As discussed above, a $(p,q)$ dyon is introduced when a $(p,q)$ string ends on a D3-brane.

\subsection{The probe approximation}

Consider a D3-brane embedded in a 10d spacetime. The probe approximation involves
neglecting the back reaction of the
brane on the geometry and the other background fields. Since the brane is a
source for one unit of flux, this requires that the background flux $N$ is large,
so that the distinction between $N$ and $N+1$ becomes negligible.
The world-volume action of a single D-brane contains a $U(1)$
field strength, $F_{\a\b}= \pa_\a A_\b - \pa_\b A_\a$. Since,
no derivatives of $F$ are included in the formula, $F$ is required to vary sufficiently
slowly so that its derivatives can be neglected. A similar restriction applies to the other
world-volume fields, as well.

Despite these approximations, the probe D3-brane action has some beautiful
exact properties: It precisely realizes the isometry of the $AdS_5 \times S^5$
background geometry as a world-volume superconformal symmetry $PSU(2,2|4)$. Only
the bosonic subgroup is taken into account in the subsequent discussion, since fermi
fields are omitted.  (A version of the complete formula with fermions is given in
\cite{Metsaev:1998hf}.) The brane action also has the duality symmetry of the
background, which is $SL(2,\IZ)$ for the D3-brane example.
Furthermore, the D3-brane world-volume action is most naturally formulated
with local symmetries, which are general coordinate invariance and a fermionic
symmetry called kappa symmetry. It is only after implementing a suitable gauge
choice that one is left with a conventional nongravitational quantum field theory
in Minkowski spacetime.

In principle, one can integrate out the massive fields of the Coulomb branch theory
exactly, thereby producing a very complicated formula in terms of the massless photon
supermultiplet only. Very schematically,
\begin{equation}
 \exp(iS_{\rm HEA}) = \int DW^+ DW^- \exp(iS)
\end{equation}
If one could do this integral exactly, which is not possible in practice, the resulting action
would capture the entire theory on the Coulomb branch, and it would be
valid at all energies. This is in contrast with the more common notion of a low-energy
effective action, which only includes the leading terms in a derivative expansion.
We have proposed to call such an exact Coulomb branch action a {\em highly effective action} (HEA).
Even though we cannot carry out such an exact computation,
in some cases we know many of the properties that the HEA should possess.
One can hope that they go a long way towards determining it.

\section{The Highly Effective Action}

\subsection{General requirements for an HEA}

An HEA should have all of the unbroken and
spontaneously broken global symmetries of the original Coulomb branch theory
with the massive $W$ fields included.
The conformal symmetry is spontaneously broken as a consequence of assigning
a vacuum expectation value (vev) to a massless scalar field that has a flat potential.
In other words, this vev spontaneously breaks spacetime symmetries
(though not Poincar\'e symmetry) as well as gauge symmetry.
An HEA should have the same duality properties as the
Coulomb branch theory containing explicit $W$ fields. The global symmetry
and duality groups are $PSU(2,2|4)$ and $SL(2,\IZ)$ in the example
considered here.

A further requirement for an HEA is that
it should have the same spectrum of supersymmetry protected (or BPS) states as
the original Coulomb branch theory. In particular, the $W^{\pm}$ supermultiplets,
which have been integrated out, should reappear in the HEA as solitons.
By contrast, this would not be expected for a low-energy effective action.
When we examine these soliton solutions explicitly, we will be led to a certain
refinement of the interpretation of this requirement.
Specifically, a complete understanding
of the BPS soliton solutions also requires knowledge of the original conformal
branch gauge theory for which the gauge symmetry and conformal symmetry are
unbroken.

Ref.~\cite{Schwarz:2013wra} conjectured that the probe D3-brane action,
in an $AdS_5 \times S^5$
background with one unit of flux ($N=1$), is precisely the HEA for the
${\cal N} =4$, 4d SYM theory with $U(2)$ gauge symmetry
on the Coulomb branch despite the fact that it is only an approximate
solution to a different problem. In fact, $N=1$ is the worst possible
case for the probe approximation. However, this shouldn't matter,
because we are solving a different problem. Still, it is helpful to keep
the 10d description in mind.

I am not certain that this conjecture is correct. On the one hand,
the formula that we will obtain seems to be too simple
to be the exact answer for such a complicated path integral. On the other
hand, we will find that it has all of the expected properties of the HEA. If
it not the conjectured HEA, then it would seem to define a new maximally
supersymmetric 4d quantum field theory, whose existence is unexpected. So either
conclusion is remarkable.

\subsection{The AdS Poincar\'e patch}

The $AdS_{p+2}$ geometry with unit radius can be described as a hypersurface
embedded in a $(p+3)$-dimensional Lorentzian space of signature $(p+1,2)$:
\begin{equation}
y \cdot y -u\phi = -1,
\end{equation}
where $ y \cdot y = -(y^0)^2 + \sum_1^p (y^i)^2$.
The equation eliminates one of the two time directions leaving a manifold
of signature $(p+1,1)$. The Poincar\'e-patch metric of radius $R$ is
\begin{equation}
ds^2 = R^2 (dy\cdot dy - du d\phi).
\end{equation}
Note that this preserves the $SO(p+1,2)$ symmetry of the embedding equation.
Defining $x^\m = y^\m/\phi$ and eliminating $u = \phi^{-1} + \phi x \cdot x$,
\begin{equation}
ds^2 = R^2(\phi^2 dx\cdot dx + \phi^{-2}dv^2).
\end{equation}
It is more customary to write the formula
in terms of a coordinate $z = \phi^{-1}$, which has dimensions of length.
However, the choice of $\phi$ is convenient for our purposes, since it will
correspond to a scalar field, also called $\phi$, with dimensions of inverse
length in the D3-brane action.

The Poincar\'e patch is not geodesically
complete. Rather there exists a different coordinate choice that describes
a geodesically complete spacetime, called global AdS,
for which the Poincar\'e patch is just
a region. The Poincar\'e patch contains a horizon, which is not present in
the global AdS metric. In this respect it is analogous to
the horizon of Rindler spacetime, whose geodesic completion is Minkowski
spacetime. The Poincar\'e patch of AdS is sufficient for our purposes.

The ten-dimensional $AdS_5 \times S^5$ metric $ds^2 = g_{MN}(x)dx^M dx^N$ is
\begin{equation}
ds^2 = R^2 \left( \phi^2 dx \cdot dx + \phi^{-2} d\phi^2 + d\O_5^2 \right)
 = R^2\left(\phi^2 dx\cdot dx + \phi^{-2} d\phi \cdot d\phi\right),
\end{equation}
where $d\O_5^2$ is the metric of a round unit-radius five-sphere, and $\phi$ is now the length of the six-vector $\phi^I$. It will be important later that $\phi$ cannot be negative.

\subsection{The D3-brane in $AdS_5 \times S^5$}

According to the AdS/CFT dictionary, the value of the radius $R$,
expressed in string units, satisfies
\begin{equation}
R^4 = 4\pi g_{\rm s} N l_{\rm s}^4 .
\end{equation}
The positive integer $N$ is the number of units of five-form flux
on the five-sphere. Also, $\int_{S^5} F_5 =2\pi N$ for an appropriate normalization,
and $N$ is the rank of the $U(N)$ gauge group of the dual CFT.

The probe D3-brane action is the sum of two terms: $S =S_1 + S_2$.
$S_1$ is a Dirac--Born--Infeld (DBI) functional of the ten embedding functions
$x^M(\s^\a)$ and a world-volume $U(1)$ gauge field $A_\b (\s^\a)$ with field strength
$F_{\a\b} = \pa_\a A_\b - \pa_\b A_\a$. $S_2$ is a Chern--Simons-like (CS) term,
which is linear in the RR fields. $S_1$ and $S_2$ also depend of the fermi fields
that we are omitting.

The general formula for the bosonic part of the DBI term is
\begin{equation}
S_1 = -T_{D3} \int \sqrt{- \det \left(G_{\a\b} + 2\pi \a' F_{\a\b}\right)}\, d^4 \s ,
\end{equation}
where $G_{\a\b}$ is the induced 4d world-volume metric
\begin{equation}
G_{\a\b} = g_{MN}(x) \pa_\a x^M \pa_\b x^N.
\end{equation}
As usual, $\a' = l_{\rm s}^2$ and the D3-brane tension is
\begin{equation}
T_{D3} = \frac{2 \pi}{g_{\rm s}(2\pi l_{\rm s})^4} .
\end{equation}
In fact, only two dimensionless combinations of parameters occur in the brane action:
\begin{equation}
R^4 T_{D3} = \frac{N}{2\pi^2}
\quad \and \quad
2\pi \a'/R^2 = \sqrt{\pi/ g_{\rm s} N}.
\end{equation}
Thus, $S_1$ only depends on the dimensionless string coupling constant $g_s$ and the
integer $N$. We shall see shortly that the $g_s$ dependence also drops out.

The nonvanishing Chern--Simons terms (aside from coefficients, which can be found in \cite{Schwarz:2013wra}) are
\begin{equation}
S_2 \sim \int C_4 + \chi \int F \wedge F.
\end{equation}
The RR four-form potential $C_4$ has a self-dual field strength $F_5 = dC_4$.
As mentioned previously,
\begin{equation}
F_5 \sim N\left( {\rm vol}(S^5) + {\rm vol} (AdS_5) \right),
\end{equation}
and $\chi$ is the value of the RR 0-form $C_0$. It is proportional to the
theta angle of the gauge theory.

$S_1$ contains a potential term $\int \phi^4 d^4x$, which should not appear in $S$,
since there should be no net force acting on the brane when $\phi >0$.
In fact, this term is canceled by the
$\int C_4$ term in $S_2$. To show this we consider the region $M$ of $AdS_5$ for
which the coordinate $\phi$ is less than the value of $\phi$ that specifies
the position of the D3-brane. Then, using Stokes' theorem,
\begin{equation}
\int_{D3} C_4 = \int_M F_5 \sim \int_0^{\phi} \int_{D3'} {\rm vol} (AdS_5)',
\end{equation}
where
\begin{equation}
{\rm vol} (AdS_5)' \sim (\phi')^3 d\phi' \wedge dx^0 \wedge dx^1 \wedge dx^2 \wedge dx^3
\end{equation}
Thus,
\begin{equation}
\int_{D3} C_4 \sim \int \phi^4 d^4x.
\end{equation}
The coefficients work perfectly.

The general coordinate invariance of the D3-brane action allows one to
fix a gauge. A convenient choice for our purposes is the static gauge
in which the four world-volume coordinates $\s^\a$ are identified with
the four Lorentzian coordinates $x^\m$, which were introduced previously,
\begin{equation}
x^\m (\s) = \d^\m_\a \s^\a .
\end{equation}
In this gauge $\phi^I (\s)$ and $A_\m (\s)$ are the only remaining bosonic
world-volume fields.
Moreover, due to the gauge choice, they become functions of $x^\m$.

The complete D3-brane action in the static gauge (aside from fermions), expressed
in terms of canonically normalized fields $\phi^I (\s)$ and $A_\m (\s)$, is
\begin{equation} \label{Scan}
S = \frac{1}{\g^2} \int  \phi^4 \left(1 - \sqrt{- \det M_{\m\n}}\right)\, d^4 x
+ \frac{1}{4} g_s \chi \int F \wedge F ,
\end{equation}
where
\begin{equation}
\g = \sqrt{N/2\pi^2}
\end{equation}
and
\begin{equation}
M_{\m\n} =  \eta_{\m\n}+ \g \frac{F_{\m\n}}{\phi^2}
+ \g^2\frac{\pa_\m \phi^I \pa_\n \phi^I}{ \phi^4} ,
\end{equation}
where $\eta_{\m\n}$ denotes the Lorentz metric.
It is an important fact that the first term in $S$ is independent of $g_s$ and $\chi$ and
that they only appear (as a product) in the coefficient of the second term.

The scale invariance of the action (\ref{Scan}) is manifest,
since all terms have dimension four and all parameters are dimensionless.
On the Coulomb branch there is a scale, the vev of the scalar field $\phi$.
However, the full conformal symmetry is realized on the action.
Only the choice of vacuum breaks the symmetry (and all choices are equivalent).
The D3-brane action contains inverse powers of the
scalar field. However, the vev of this field ensures that these are not singular.
Individual terms in the action can be arbitrarily complicated
and still end up with dimension four by including an
appropriate (inverse) power of $\phi$.

Regarded as an HEA, (\ref{Scan}) should encode all the quantum effects
that arise from integrating out the $W^\pm$ supermultiplets. On the other hand, (\ref{Scan}) should
have its own loop expansion (or path integral) to take account of the quantum effects of the
massless supermultiplet. The loop expansion is expected to be free of UV divergences, because
of the (spontaneously broken) conformal symmetry. There are IR divergences, but
they can be treated by standard methods. Rescaling all fields in (\ref{Scan})
by $\g$ brings the action to a form in which all the $N$ dependence
appears as an overall factor of $N$, so that $S(N) = NS(1)$.  This shows that the loop expansion of this theory is a $1/N$ expansion.

\subsection{S duality}

The $SL(2, \IZ)$ transformation $\t \to -1/\t$ duality is accompanied by an electric-magnetic
transformation in the S-duality transformation of the gauge theory. This invariance
has not been proved in the nonabelian formulation with $W$ fields, but it was
proved for the abelian D3-brane action in \cite{Schwarz:2013wra}.
Recall that $\t = \t_1 + i \t_2 = \chi + {i}/{g_s}$, but that the action only depends
on $g_s \chi = \t_1/\t_2$. When $\t \to -1/\t$,  $g_s \chi \to - g_s \chi$. To show
invariance under this sign change, we must also examine the transformation
that exchanges electric and magnetic fields.

The procedure is standard. It involves replacing the $U(1)$ gauge field $A_\m$ by
a new one such that the Bianchi
identity
\begin{equation}
\pa_{[\m} F_{\n\r]} =0
\end{equation}
becomes a field equation, and the field equation
\begin{equation}
\pa_{\m}\left( \frac{\pa {\cal L}}{\pa F_{\m\n}} \right) =0
\end{equation}
becomes a Bianchi identity.
The nontrivial fact, which was proved in \cite{Schwarz:2013wra}, is that the new action, obtained by this field replacement together with the sign change
$g_s \chi \to - g_s \chi$, is identical to the original one.

\subsection{Summary and discussion}

We have conjectured that the world-volume action of a probe D3-brane in
an $AdS_5 \times S^5$ background of type IIB superstring theory,
with one unit of flux ($N$=1), can be reinterpreted as the HEA of $U(2)$ ${\cal N} =4$
SYM theory on the Coulomb branch. An explicit formula for the bosonic part of the action
has been presented. It is likely that there is a generalization in which
the formula with $N>1$ plays a role in the
Coulomb branch decomposition $U(N+1) \to U(N) \times U(1)$. However, when $N>1$
certain issues still need to be clarified.

The evidence presented so far for the conjecture that the
probe D3-brane action is the desired HEA is that the action incorporates all
of the required symmetries and dualities: $PSU(2,2|4)$ superconformal
symmetry (when fermions are included) and $SL(2,\IZ)$ duality.
The next section will describe BPS soliton solutions of this action.
Their properties will give further support to the conjecture. They will also
lead us to refine the conjecture somewhat.

\section{BPS Soliton Solutions}

The previous section described the construction of the bosonic part of
a world-volume action for a probe D3-brane in $AdS_{5}\times S^5$.
We also discussed ${\cal N} =4$ SYM theory, with gauge group $U(N+1)$,
on the Coulomb branch. A suitable scalar field vev breaks the gauge symmetry so that
\begin{equation}
U(N+1) \to U(1) \times U(N).
\end{equation}
The $2N$ supermultiplets corresponding to the broken symmetries
acquire mass $gv$. The probe D3-brane action was
conjectured to be the HEA for the $U(1)$ factor.

In this section we will derive classical
supersymmetric soliton solutions of the action (\ref{Scan}). Even though they are
classical, they are supposed to take account of all quantum effects due
to the massive fields that have been integrated out. Furthermore, they will
turn out be BPS, which implies that they are protected from additional
quantum corrections. The soliton solutions will be independent of
$N$ (or $\g$), since $N$ is an overall factor of the action written in
terms of rescaled fields. Thus, this parameter drops out of the classical field
equations.

For the purpose of constructing the soliton solutions, it is sufficient
to consider a D3-brane that is localized at a fixed position on the $S^5$.
In this case, the $S^5$ coordinates do not contribute to the D3-brane action.
The radial position in the $AdS_5$ space is encoded in the
nonnegative scalar field $\phi$.
The dependence of the action on this field and the $U(1)$ gauge field are all that
are required in this section. The fermions are not required, since they
vanish for these solutions.

\subsection{Solitons}

Let us begin by looking for spherically symmetrical static solutions, centered at $r=0$,
for which the action Eq.~(\ref{Scan}) is stationary.
We require that $\vec E$ and $\vec B$ (the electric and magnetic parts of $F_{\m\n}$) only
have radial components, denoted $E$ and $B$. Also,
$E$, $B$, and $\phi$ are functions of the radial coordinate $r$ only.
It then follows that
$$
-\det (M_{\m\n})
=-\det \left(\eta_{\m\n} + \g^2\frac{\pa_\m \phi \pa_\n \phi}{\phi^4}
+ \g \frac{F_{\m\n}}{\phi^2}\right)
$$
\begin{equation}
=  \left(1  + \g^2 \frac{(\phi')^2 - E^2}{\phi^4}\right)
\left( 1 + \g^2 \frac{B^2}{\phi^4}\right).
\end{equation}
This results in the Lagrangian density
\begin{equation}
{\cal L} = \frac{1}{\g^2}  \phi^4\left(1 -
\sqrt{\left( 1 + \g^2\frac{(\phi')^2 - E^2}{\phi^4}\right)
\left(1 + \g^2 \frac{ B^2}{\phi^4}\right)} \, \right) + g_s \chi B E.
\end{equation}
The equation of motion for $A_0$ is $ \frac{\pa}{\pa r} (r^2D) =0,$ where
\begin{equation}
D = \frac{\pa {\cal L}}{\pa E}
= E \sqrt{\frac{1+\g^2 B^2/\phi^4}{1 + \g^2 [(\phi')^2 - E^2]/\phi^4}} +g_s \chi B.
\end{equation}
For a soliton centered at $r=0$, with $p$ units of electric charge $g$ and $q$ units
of magnetic charge $g_m$, where $g_m = 4\pi/g$, we have
\begin{equation}
D =  \frac{p g}{4\pi r^2} \quad \and \quad B = -\frac{q g_m}{4\pi r^2}.
\end{equation}

The mass of the soliton is given by the Hamiltonian.
$ H = \int {\cal H} d^3x  = 4\pi \int {\cal H} r^2 dr $. The Hamiltonian
density is ${\cal H} = DE - {\cal L}$. Eliminating $E$ in favor of $D$ gives
\begin{equation} \label{hamden}
 {\cal H} = \frac{1}{\g^2} \left(\sqrt {(\phi^4 + \g^2 (\phi')^2)
(\phi^4 + \g^2 X^2)} -\phi^4 \right),
\end{equation}
where
\begin{equation}
X =  \sqrt{\tilde D^2 + B^2} = Q/r^2.
\end{equation}
and
\begin{equation}
\tilde D =D -g_s \chi B.
\end{equation}
Thus, $ \tilde D^2 + B^2 = Q^2/r^4, $ where
\begin{equation}
Q = \frac{g}{4\pi} |p + q\t|.
\end{equation}
As before,
\begin{equation}
\t = \chi + i/g_s = \frac{\th}{2\pi} + i \frac{4\pi}{g^2}.
\end{equation}

We want to find functions $\phi(r)$ that give BPS extrema of $H$
with the boundary condition $\phi \to v$ as $r \to \infty$.
The BPS condition turns out to require that the two factors inside the
square root in Eq.~(\ref{hamden}) are equal, which implies that ${\cal H} = (\phi')^2$.
The proof goes as follows. One first writes the formula for ${\cal H}$
in the form
\begin{equation}
(\g^2 {\cal H} + \phi^4)^2 = ( \g^2 X |\phi'|
+ \phi^4)^2 + \g^2 \phi^4 (X - |\phi'|)^2.
\end{equation}
Thus,
\begin{equation}
 (\g^2 {\cal H} + \phi^4)^2 \geq ( \g^2 X |\phi'| + \phi^4)^2,
\end{equation}
which implies ${\cal H} \geq X|\phi'|$. Saturation of the BPS bound
is achieved for $|\phi'| =X$ and then
\begin{equation}
{\cal H} = X^2 = (\phi')^2 = Q^2/r^4.
\end{equation}

The equation $(\phi')^2 = Q^2/r^4$, together with the boundary condition
$\phi \to v$ as $r \to \infty$, has two BPS solutions
\begin{equation}
\phi_{\pm}(r) = v \pm Q/r ,
\end{equation}
where $Q = \frac{g}{4\pi} |p + q\t|$.
The $\phi_+$ solution is similar to the flat-space case
studied by Callan and Maldacena \cite{Callan:1997kz}. It describes
a funnel-shaped protrusion of the D3-brane extending to the
boundary of AdS at $\phi=+\infty$. This solution gives
infinite mass (proportional to $\int dr/r^2$), and thus
it is not the solution we are seeking.

The $\phi_-$ solution is different. $\phi =0 $ corresponds
to the horizon of the Poincar\'e patch of $AdS_5$. Thus, since $\phi$
is nonnegative, the $\phi_-$ solution must be cut off at
\begin{equation}
r_0 = \frac{Q}{v}.
\end{equation}
Then the masses of BPS solitons are given by
\begin{equation}
M = 4\pi \int_{r_0}^{\infty} {\cal H} r^2 dr
= \frac{4\pi Q^2}{r_0} = 4\pi v^2 r_0 = v g |p+q\t|,
\end{equation}
exactly as was expected. We have
obtained a complete infinite $SL(2,\IZ)$ multiplet, which includes the $W$
particles that were integrated out, as well as monopoles and dyons.

The charge of the $\phi_-$ solution is uniformly spread on the sphere $r=r_0$,
which we call a {\em soliton bubble}.
The interior of the bubble should not contribute to the mass of the soliton.
So, how should we think about the interior of the bubble in the QFT?
The only sensible interpretation is that the gauge theory
is in the ground state of the {\em conformal phase} of $U(N+1)$
inside the sphere. This implies that the bubble is a {\em phase boundary}.
This interpretation has the advantage that the parameter $\t$ is required to
describe the $U(N+1)$ theory in the conformal phase. This would explain how
the soliton solutions know what the values of $g$ and $\th$ are. The $U(1)$
action does not contain all the required information. One also needs to know the
nonabelian theory in the conformal phase.

\subsection{Comparison with the BPS 't Hooft--Polyakov monopole}

Let us compare the monopole solution of the D3-brane action that we have
obtained with the corresponding BPS 't Hooft--Polyakov monopole solution.
The single monopole solution of the
nonabelian $SU(2)$ gauge theory on the Coulomb branch (for $\th =0$)
has a triplet of scalar fields $\phi^a$ whose internal symmetry is aligned with
the spatial directions as follows\footnote{The ${\cal N}=4$ theory has
six such triplets, but only one of them is utilized in the construction. This
choice corresponds to the choice of a point on the five-sphere in the D3-brane
construction.}:
\begin{equation}
\phi^a (\vec x) = \frac{x^a}{r} \phi(y)
\end{equation}
with
\begin{equation}\label{tHP}
\phi(y) = v ({\rm coth} y - 1/y),
\end{equation}
where $y = M_W r$ and $M_W =gv$.
$\phi(y)$ is strictly positive for $y>0$, and $\phi^a (\vec x)$
is nonsingular at the origin. Thus, there is no sign of a soliton bubble
in the nonabelian description. Both constructions give the correct mass and
charge for the monopole, but the D3-brane solution gives a soliton bubble
whereas the nonabelian solution does not. So, which formula more accurately
describes what is happening?

Equation~(\ref{tHP}) differs from
\begin{equation}
\phi_-(r) = v(1 - 1/y),
\end{equation}
the D3-brane theory result for the monopole,
by a series of terms of the form $\exp(-2n M_W r)$, where $n$ is a positive integer.
In the context of the ${\cal N} =4$ theory, the effect of integrating out the fields
of mass $M_W$ should be to cancel these exponential terms for $y>1$ and
to give $\phi =0$ for $y<1$. After all, the $U(1)$ HEA is supposed to
incorporate all of the contributions due to $W$ loops, and these exponentials are
a plausible form for those contributions.  Hence, we conclude
that the bubble is real and that the D3-brane solution gives a more accurate description
of what is happening than the usual classical solution (\ref{tHP}) of the nonabelian theory.

\subsection{Black hole analogy}

We have found a universal formula relating the mass and radius of BPS soliton
bubbles: $ M = 4\pi v^2 r_0 $, which is valid for all $(p,q)$.
For comparison, the radius of the horizon of a 4d extremal Reissner--Nordstrom
asymptotically Lorentzian black hole in four dimensions is $r_0 = MG$, where $G$
is Newton's constant, for all $(p,q)$.
In the latter case the charge $Q$ should be large for the classical analysis to
be valid. Thus, the relation between mass and radius is the same in both cases,
with $(4\pi v^2)^{-1}$  the analog of Newton's constant. The BPS condition
ensures that the analogy extends to the relation between mass and charge.

This analogy is rather surprising, because the D3-brane theory is a nongravitational
theory in 4d Minkowski spacetime. If one tries to pursue this analogy, there is
a natural question: Does the Bekenstein--Hawking entropy of the black hole, which is
proportional to $Q^2$ (for large $Q$), have an analog for the solitons of the D3-brane theory?
For example, is there an entanglement entropy between the inside and outside of the soliton
bubble with this value. If the solitons can be shown to have a well-defined entropy of this sort
(for large $Q$), then one may be tempted to take the black-hole analogy seriously.

Even though the D3-brane theory is defined on a 4d Lorentzian spacetime, we know that
the field $\phi$ can be interpreted as a radial coordinate in $AdS_5$. From this point
of view the soliton solutions have a nontrivial geometry induced from their embedding
in $AdS_5$. From the 5d (or 10d) viewpoint, the bubble is on the horizon of the
Poincar\'e patch of $AdS_5$, where it intersects the boundary of global AdS.
This fact may be useful for understanding the origin of the black-hole analogy.

\subsection{Multi-soliton solutions}

It is easy to derive the generalization of $\phi_-(r)$ to the
case of $n$ solitons of equal charge.\footnote{Pairs of solitons with different charges are
mutually nonlocal and therefore difficult to describe.}
Since supersymmetry ensures that the forces between them should cancel
when they are at rest, their centers can be
at arbitrary positions $\vec x_k$, $ k = 1,2,\ldots,n$.
Since $\phi$ satisfies Laplace's equation, the solutions can
be superposed. The obvious guess, which is easy to verify, is
\begin{equation}
\phi (\vec x) = v - Q \sum_{k=1}^n \frac{1}{|\vec x - \vec x_k|}.
\end{equation}
The surfaces of the bubbles, which are no longer spheres,
are given by $\phi(\vec x) =0$. The fields
$\vec D$ and $\vec B$ are then proportional to $ \vec\nabla \phi$,
with coefficients determined by the charges. Their values at $\phi =0$
determine the charge densities on the bubble surface. This is much simpler
than the usual multi-monopole analysis, which involves nonlinear equations!
It should also be more accurate.

\subsection{Previous related work}

Soliton bubbles like those found here have appeared in the literature previously.
We will briefly discuss the examples that we are aware of. There may be others.

Using attractor flow equations \cite{Ferrara:1995ih},
Denef \cite{Denef:1998sv}, \cite{Denef:2000nb}
found similar structures to our soliton
bubbles in the context of supergravity solutions in which a D3-brane wraps a cycle
of a Calabi--Yau manifold that vanishes at a conifold point, where the central
charge modulus is zero.

Gauntlett {\it et al.} \cite{Gauntlett:1999xz} studied soliton solutions
of a probe D3-brane in an asymptotically flat black D3-brane supergravity
background. This problem is closely related to the one we have considered,
since this geometry has $AdS_5 \times S^5$
as its near-horizon limit. They identified half-BPS solutions,
like those found here, ``in which a point charge is replaced by a perfectly conducting
spherical shell.'' In \cite{deMelloKoch:1999ui} the authors examined the DBI action
of a D3-brane probe in F theory. They constructed a monopole solution containing
a soliton bubble that coincides with a 7-brane.
The formation of soliton bubbles may also be related to the {\em enhan\c{c}on} mechanism
in \cite{Johnson:1999qt}. This mechanism circumvents the appearance of a class of
naked singularities, known as {\em repulsons}.

In \cite{Popescu:2001rf}, Popescu and Shapere studied the low-energy effective action
of ${\cal N} =2 $ $SU(2)$ gauge theory without additional matter in the Coulomb phase.
The Seiberg--Witten {\em low-energy} effective action  \cite{Seiberg:1994rs}
should only be valid below the mass of $W$ bosons.
However, the theory has a BPS monopole and a BPS dyon that become massless
at points in the moduli space of vacua. Therefore, these particles should be obtainable
as soliton solutions of the Seiberg--Witten action. Popescu and Shapere constructed these
solutions and discovered that they exhibit a spherical shell of charge, just like what we have found.

There is some interesting related evidence for soliton bubbles in a nonsupersymmetric
field theory context \cite{Bolognesi:2005rk}--\cite{Manton:2011vm}. By considering
multi-monopole solutions of large
magnetic charge in the 4d $SU(2)$ gauge theory with adjoint scalars on the
Coulomb branch, Bolognesi deduced the existence of ``magnetic bags'' with properties
that are very close to those of the soliton bubbles obtained here.
Bolognesi's magnetic bags do not have sharply defined surfaces, though they become
sharp in the limit of large charge.  Bolognesi also pointed out the
analogy to black holes \cite{Bolognesi:2010xt}.

\section{Conclusion}

The action of a probe D3-brane
in $AdS_5 \times S^5$ is a candidate for the HEA for a $U(1)$
factor of ${\cal N} =4$ SYM theory on the Coulomb branch.
It incorporates all the required symmetries
and dualities, and it gives the expected BPS soliton solutions.
Even so, it might only be an approximation to the true HEA that is sufficient for computing
susy-protected quantities. This would also be noteworthy, since then it would be a
candidate for a new interacting UV finite abelian ${\cal N} =4$ quantum field theory in 4d.
The existence of such a theory is unexpected. In either case, it is important to
settle this question.

There is much more that remains to be explored. We need to
understand the extent to which symmetry and other general considerations
determine the HEA and whether the world-volume theory
of a probe D3-brane should give this HEA. We should construct other analogous
$p$-brane actions and explore their BPS soliton solutions. (BPS soliton solutions
of the M5-brane example are discussed briefly in \cite{Schwarz:2014rxa}.) It should be
illuminating to explore tree-approximation scattering amplitudes of the
D3-brane theory. It seems likely that they will exhibit beautiful properties,
maybe even a (spontaneously broken) Yangian symmetry. Finally, we would like to
generalize the analysis to higher-rank gauge theories on the Coulomb branch,
which have multiple abelian factors.

\section*{Acknowledgment}

The author wishes to acknowledges discussions and communications with
Hee-Joong Chung, Frank Ferrari, Abhijit Gadde, Jerome Gauntlett, Sergei Gukov,
Nicholas Hunter-Jones, Elias Kiritsis, Arthur Lipstein, Nick Manton,
Hirosi Ooguri, Jaemo Park, Nati Seiberg, Savdeep Sethi, Yuji Tachikawa,
David Tong, and Wenbin Yan.

This work was supported in part by DOE Grant \# DE-SC0011632.
The author acknowledges the hospitality of the Aspen Center
for Physics, where he began working on this project in the summer of 2013,
and where he wrote this manuscript in the summer of 2014.
The ACP is supported by the National Science Foundation Grant No. PHY-1066293.

\end{document}